\documentclass[twocolumn]{webofc}
\usepackage[varg]{txfonts}
\usepackage{hyperref}
\usepackage{cleveref}
\usepackage{bm}

\title{CORSIKA 8 -- Towards a modern framework for the simulation of extensive air showers}
\author{\firstname{Maximilian} \lastname{Reininghaus}\inst{1,2}
 \and \firstname{Ralf} \lastname{Ulrich}\inst{1}
 on behalf of the CORSIKA~8 Project
}

\hypersetup{
  pdftitle={CORSIKA 8 -- Towards a modern framework for the simulation of extensive air showers},
  pdfauthor={Maximilian Reininghaus, Ralf Ulrich},
  pdflang=en
}

\institute{Institut für Kernphysik, Karlsruher Institut für Technologie (KIT), Karlsruhe, Germany
\and
 Institut für Experimentelle Teilchenphysik, Karlsruher Institut für Technologie (KIT), Karlsruhe, Germany}

\abstract{Current and future challenges in astroparticle physics require
  novel simulation tools to achieve higher precision and more
  flexibility.  For three decades the FORTRAN version of CORSIKA
  served the community in an excellent way. However, the effort to
  maintain and further develop this complex package is getting
  increasingly difficult. To overcome existing limitations, and
  designed as a very open platform for all particle cascade simulations
  in astroparticle physics, we are developing CORSIKA~8 based on modern
  C++ and Python concepts. Here, we give a brief status report of
  the project.}

\begin{document}
\maketitle

\section{Introduction}
CORSIKA \cite{Capdevielle:1992qw,Heck:1998vt} is the most widely used,
actively maintained code for Monte Carlo air shower simulation currently available. In spite
of its development having started almost 30 years ago \cite{Gils:1989vi},
it is still frequently extended and improved, with major updates released roughly once per year
consisting mostly of improvements in the various interaction models shipped with CORSIKA.
Completely new features, however, are developed only rarely
nowadays, also due to the complexity of the code posing a major obstacle to
their implementation: Originally written and optimized 
to be used only in simulations for the KASCADE experiment~\cite{Klages:1996zb}, and
therefore designed to meet the corresponding requirements, it was not intended
to serve as the general purpose tool into which it has evolved. Its monolithic
structure makes modifications or extensions of the code very difficult.
Besides that, CORSIKA is written in FORTRAN~77, which can no longer be considered
the lingua franca within the domains of high energy and astroparticle
physics, and suffers from a number of restrictions, e.g.\ the lack of dynamic memory
or object orientation, and is therefore unattractive to learn, causing a
lack of qualified and motivated contributors.

Although the C++ add-ons \textit{COAST}~\cite{Ulrich:2006} or recently
\textit{dynstack}~\cite{Baack:2016} help to remedy parts of these
issues to a certain degree, there are still many wishes by users for extensions that
simply cannot be accommodated for with reasonable effort. It is
clearly a disadvantage to wrap modern extensions around the existing
"dinosource"~\cite{Zwart:2018} code in comparison to fundamentally re-designing the
whole framework in a consistent way. 

For that reason, we reached the decision that the time has come to start
a project to develop a next-generation code, with the focus on the aspects
modularity, flexibility, ease of use and extensibility, efficiency, and reliability
from the beginning.
Of course, a key element of the new project is to keep the expertise gained and include
all the lessons learned from the last decades. 
While the name is chosen to abide, the distinction between
the legacy and next-generation CORSIKA is made through the version number,
initially CORSIKA~8 for the latter. We consider CORSIKA~8 to be more of a framework
for simulating particle cascades rather than an air-shower-only tool, therefore
extending the applicability to wider domains of research.

To a large extent, our goals and plans are outlined in ref.~\cite{Engel:2018akg}
and their implementation is currently ongoing work. Here, we present an
overview of the most important aspects of the design.

\section{Building blocks}
\begin{figure*}[tb]
  \centering
  \includegraphics[width=.6\textwidth]{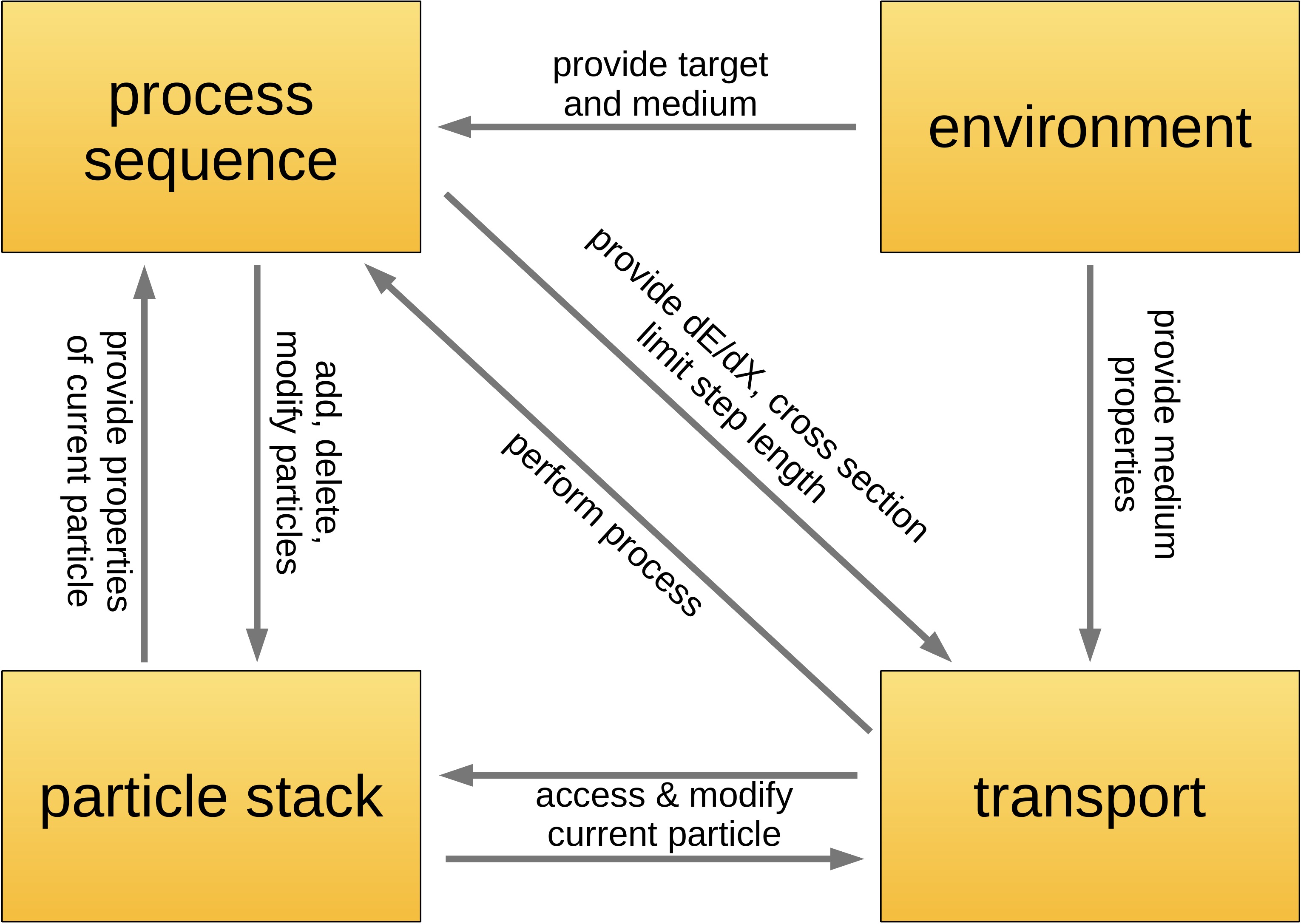}
  \caption{Main building blocks of CORSIKA~8. The relationships and
    dependencies between the blocks are shown as arrows. Each of these
    major blocks is by itself a modular system of
    algorithms. Basically all functionality can be replaced by alternative modules/implementations in a very
    straightforward way. This also means future extensions can be included easily. }\label{fig:building_blocks}
\end{figure*}

CORSIKA~8 is developed using modern C++ accompanied by Python tools.
The main building blocks of CORSIKA~8 are displayed in \cref{fig:building_blocks}
together with their relations to each other.

\subsection{Particle stack}
The \emph{particle stack} contains the particles in memory which are currently
in the course of being propagated. In its most basic incarnation the stack provides
access to the particles' four-momenta, four-positions, and particle codes, but
an easy extension of additional properties like statistical weight,
as necessary e.g.\ for thinning algorithms, is straightforward. It is envisaged
to provide optional access to the history of the particle offering a much deeper
insight into its "ancestor" generations than it is currently possible with the corresponding
feature~\cite{Heck:2009zza} of legacy CORSIKA. The particle stack
is read from and written to by the \emph{process sequence}, as well as the
\emph{transport} procedure. 

\subsection{Process sequence}
\begin{figure*}[tb]
  \centering
  \includegraphics[width=.6\textwidth]{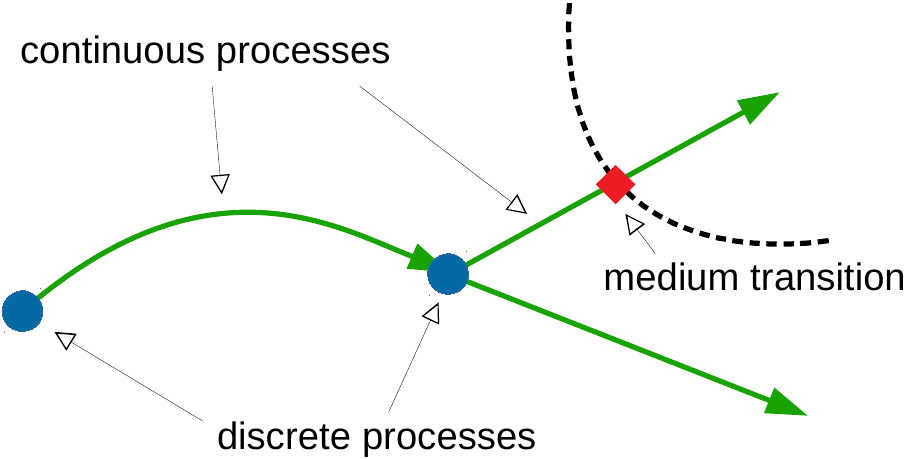}
  \caption{Discrete (blue dots) and continuous (green lines) processes
    during particle transport. The sampled random locations of
    discrete processes, which can be interactions, decays or boundary
    crossings (red square), determine the regime for continuous processes.}\label{fig:processes}
\end{figure*}

The \emph{process sequence} represents the physical processes modeled in the
simulation and is composed of all the physics modules which
the user chooses to enable (e.g.\ hadronic and electromagnetic interaction models,
emission of Cherenkov light or radio). All these modules must conform to the same
interfaces and are treated on the same level. We distinguish mainly between \emph{continuous processes} and \emph{discrete
processes} (see \cref{fig:processes}). While the first ones are meant to model
effects which happen along the trajectory between two steps of the particle
(like energy losses or Cherenkov light emission), the second type of processes
typically represents interactions and decays. Furthermore, a special class of
processes is reserved for the cases in which a particle transits the boundary
between two media.

Discrete processes need to provide interaction lengths or decay times as
a function of the particle currently being propagated (implicitly having
access to information about the local environment of the particle through
its location). In addition, in case
one of the discrete processes is then chosen to be in fact performed, that specific process can then modify
the particle stack, typically by deleting the projectile from the stack and placing
new secondaries of the interaction onto the stack.

Instead of providing an interaction length, continuous processes
are individually required to provide a maximum step-size. This is useful, inter
alia, to limit energy losses to a tolerable amount between two steps, which
would otherwise invalidate the interaction length calculated previously
using the particle energy at the beginning of the step. Continuous processes
are provided with the current particle together with the trajectory up to
its endpoint determined from a number of criteria (see below) including the
above-mentioned limited step-size. In contrast to the random nature of discrete
processes, they are always performed.

\subsection{Environment}
One of the most prominent features of CORSIKA~8 is the flexible definition
of the medium and its properties in which the particles propagate. In particular,
it will be possible to simulate not only pure air showers but also showers
penetrating the ground and propagating further through ice, water, rock,
or other media. A key premise of this endeavour is the ability to compose
the \emph{environment} out of several different (sub-)volumina with different
physical properties. In this regard, CORSIKA~8 follows similar concepts
as the well-known Geant4 toolkit~\cite{Agostinelli:2002hh,Allison:2006ve,Allison:2016lfl},
with the major difference perhaps being that we do not limit ourselves to
homogeneous media.

\begin{figure*}[tb]
  \centering
  \begin{minipage}{.58\textwidth}
    \centering
    \includegraphics[width=\textwidth]{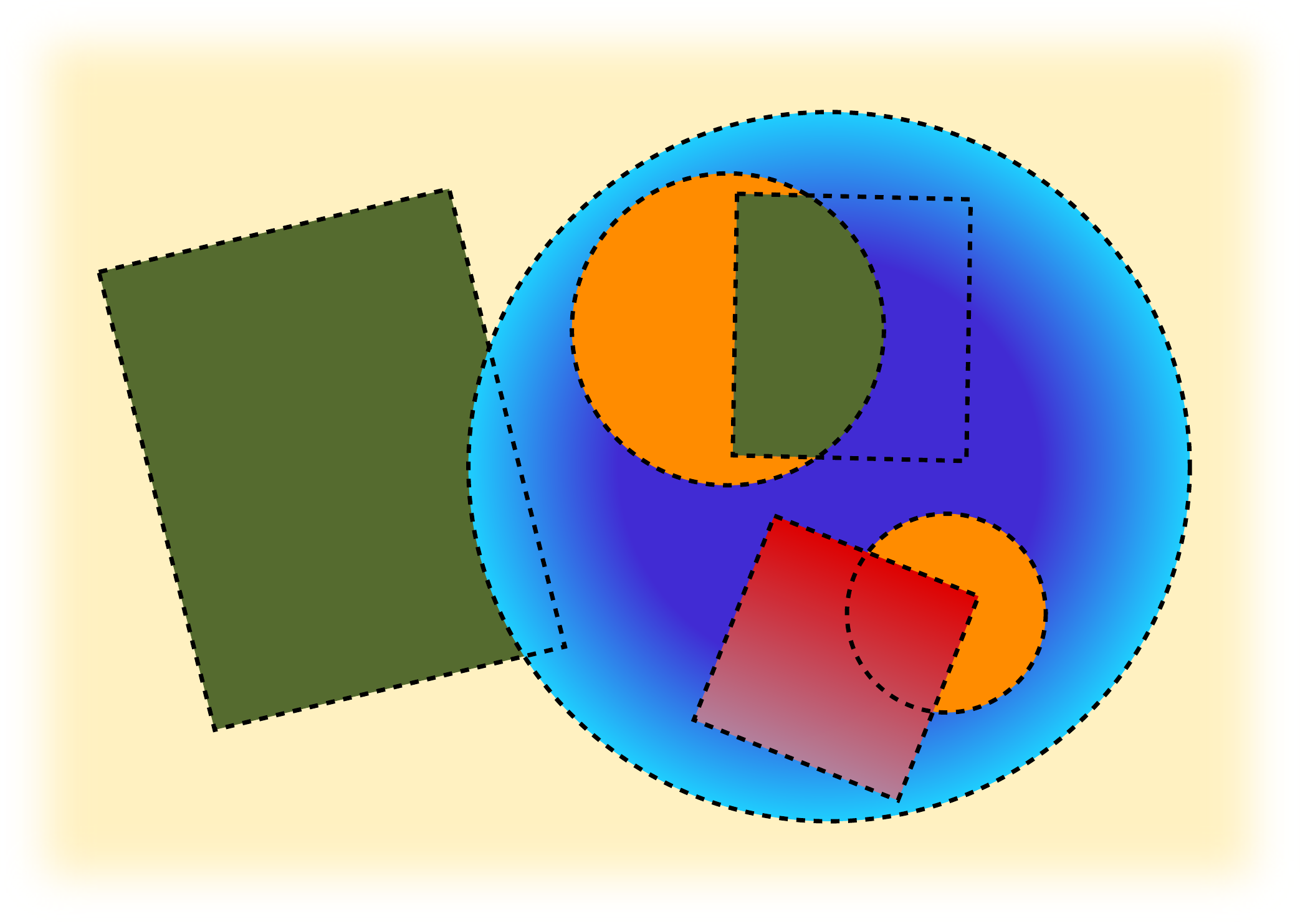}
  \end{minipage}%
  \begin{minipage}{.42\textwidth}
    \centering
    \includegraphics[width=\textwidth]{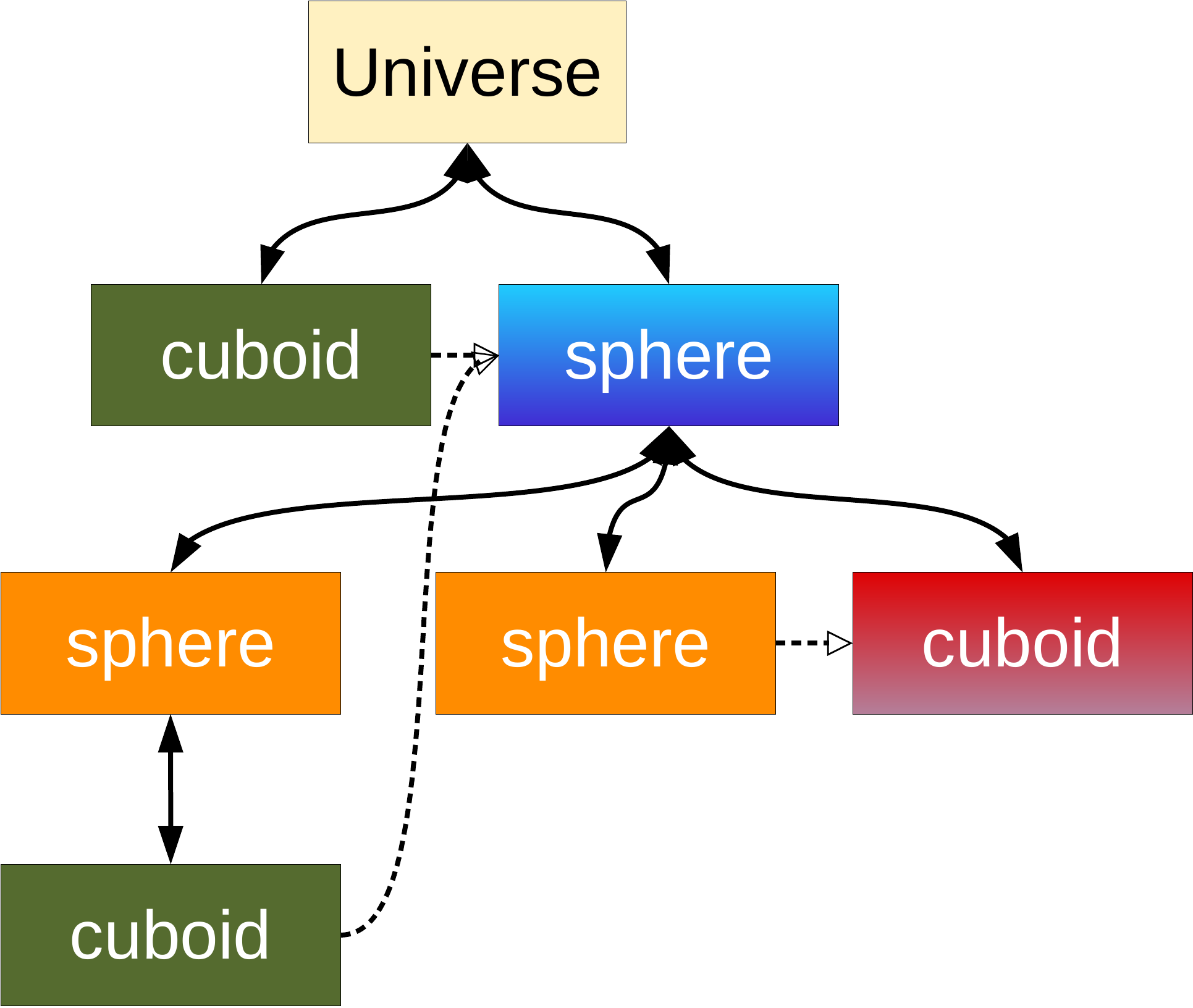}
  \end{minipage}
  \caption{An example environment composed of different volumes
	with different physical properties indicated by color (left). In the
	implementation, these are assembled in a tree structure (right).}
  \label{fig:environment}
\end{figure*}

\Cref{fig:environment} illustrates this idea together
with a sketch of the current implementation. We provide very simple volumina,
in the beginning only spheres and cuboids, which the user has to furnish
with models of its physical properties (in the figure symbolized
by the different colors) and then assemble them into the \emph{volume tree}.
The structure of the volume tree represents geometrical containment, i.e.,
volumes fully contained by a bigger volume are child nodes of the bigger volume.
The root node is always the \emph{Universe} volume which is equivalent to a
sphere with infinite radius. By relaxing the condition of full containment, 
it is possible to cut the child volume along the boundary of its parent.
Furthermore, it is necessary to treat cases of overlapping
nodes specially in order to avoid ambiguities. We achieve this by having references to other
volumes that are to be excluded from a given volume node, in the figure indicated by
the dashed arrows. The tree structure allows relatively fast queries of
which actual volume contains a given point.

As a second element of the environment, it is foreseen in the design to conveniently
change and extend the number of physical properties represented in the medium
model. As a first step, we provide interfaces for querying mass density, fractional
elementary composition, and magnetic field only. As soon as physics modules e.g.\ for
Cherenkov or radio emission requiring the index of refraction are added to
CORSIKA~8, this additional property can then easily be included. For runs
without these processes enabled, however, a definition will not be required.

\subsection{Transport}
At the heart of CORSIKA~8 lies the transport code which, making use of the
aforementioned building blocks, propagates the particles one by one, most
likely producing secondaries, until the simulation finishes -- by construction
as soon as no particles are left.

The first
step consists of proposing a trajectory, starting at the current position of
the particle and initially extending to the next point of intersection with
a volume boundary. We currently restrict ourselves to linear trajectories
since in that case the calculations of intersections with spheres and cuboids
reduce to solving polynomial equations of at most second order. For helices,
which would be a natural choice for trajectories of charged particles in
slowly varying magnetic fields, already the calculation of intersections
with planes requires a non-trivial numerical treatment~\cite{Nievergelt:1996}.

As second step, the maximum step-size is then further limited by, depending on the
environment, up to two conditions concerning the numerical accuracy of the
procedure. One limit regards the accuracy of integrating the equations of
motion in the magnetic field to make sure that the trajectory will not deviate
too much from the true, helix-like solution. This is obviously superfluous in
the absence of a magnetic field or for neutral particles. The second limit
pertains to the calculation of grammage $X$ along the trajectory within the
medium with a given density distribution $\varrho(\bm{x})$, i.e.
\begin{equation}
X = \int\limits_{\mathrm{trajectory}} \varrho(\bm x) \, \mathrm{d}s.
\end{equation}
This can be done analytically exact only for very specific density distributions,
e.g.\ a homogeneous one. For the general case, one needs to deal with either
numerical integration or approximations: a suitable approach can be
to approximate the density distribution $\varrho$ in the vicinity of the starting
point $\bm{x}_0$ of the trajectory using Taylor's expansion, say to second order,
\begin{align}
  \begin{split}
    \varrho(\bm x_0 + \delta \bm{x}) &= \varrho(\bm{x}_0) + 
    \left(\left.\bm{\nabla}\right\rvert_{\bm{x}_0} \varrho\right) \cdot
    \delta \bm{x} \\
     &+\frac12 \delta \bm{x}^{\mathrm T} \left.H\right\rvert_{\bm{x}_0} \delta\bm{x}
     + \mathcal{O}\left(\delta x^3\right),
  \end{split}
\end{align}
where $H$ denotes the Hesse matrix of $\varrho$ and $\delta \bm{x}$ is a small
piece along the trajectory. Then, the problem reduces to the integration of
a polynomial and $\delta \bm{x}$ would be limited to a certain length by requiring
the estimated error of the approximation to be smaller than a given value.

The next step consists of randomly sampling the next location of the discrete
processes. Decay points are sampled from an exponential distribution
in length, whereas for interaction points the exponentially distributed variable
is grammage. To determine the location of the interaction, grammage needs to be converted
back to length. Hence, an accurate conversion between these two variables is required.
Afterwards, continuous processes are performed
along the trajectory up to either its endpoint given by the limiting conditions
described above, or the interaction point of the closest discrete process,
which will be performed subsequently.

\section{Conclusions and outlook}
The development of CORSIKA~8 is in a very active stage. The project is
completely open to input from the community. Any participation and
collaboration will lead to a better tool for astroparticle physics for
the next decades.

We are committed to provide a reliable, stable,
accurate and flexible framework. The design as a framework, in contrast
to a single-purpose program, makes it clear that the range of future applications
could be far beyond just simulating extensive air shower cascades. It
is also up to the community to define what is needed and what is
scientifically useful. The inherent complexity of particle shower
development in materials requires a very careful validation of each
ingredient and input model, best with dedicated data. We aim to
facilitate a better understanding and study of the relationship
between these fundamental ingredients and the final physics
observables.

The first intermediate development snapshots of CORSIKA~8 are already
available on our gitlab server~\cite{gitlab} and can be obtained freely from there. We
welcome any comments or, even better, participation/discussion in
further developing this project. It is our plan to have a first
version suitable for limited and specialized physics studies available already
in 2019.

\section*{Acknowledgements}
\begin{acknowledgement}
M.R.\ acknowledges support by the DFG-funded Doctoral School ``Karlsruhe
School of Elementary and Astroparticle Physics: Science and Technology''.
\end{acknowledgement}

\bibliography{literature.bib}

\end{document}